# On Designing a Generic Framework for Cloud-based Big Data Analytics

Samiya Khan
Ph.D. Student
Department of Computer Science
Jamia Millia Islamia, New Delhi, India
samiyashaukat@yahoo.com

Mansaf Alam
Supervisor
Department of Computer Science
Jamia Millia Islamia, New Delhi, India
malam2@jmi.ac.in

## ABSTRACT
Big data analytics has gathered immense research attention lately because of its ability to harness useful information from heaps of data. Cloud computing has been adjudged as one of the best infrastructural solutions for implementation of big data analytics. This research paper proposes a five-layer model for cloud-based big data analytics that uses dew computing and edge computing concepts. Besides this, the paper also presents an approach for creation of custom big data stack by selecting technologies on the basis of identified data and computing models for the application.

## 1 INTRODUCTION
The rise of big data and the realization of its usefulness for applications have steered active research in this field. Big data analytics has found diverse applications in varied domains. Recently fields like manufacturing [1], geoscience [2], healthcare [3] and smart cities [4], have also found big data analytics viable and productive. Besides this, analytics have found applications in fields like journalism [5], agriculture [6] and education [7] as well.

With the increasing popularity of cloud-based technologies for big data analytics, application development has become simpler and accessible to a wider range of audience. This research paper proposes a five-layer model for cloud-based big data analytics. This is a novel concept model that makes use of edge computing and dew computing fundamentals to map the communication processes involved in big data systems. The proposed approach also presents a generic technology selection criteria based on the identification of data and computing models for an application.

The rest of the paper is organized as follows: Section 2 gives a brief description of the big data lifecycle to create a background of the processes that need to be catered by the big data stack. Section 3 introduces the proposed model and the function of each layer. It also explains the proposed model with reference to the big data lifecycle.

The mapping of the different layers of the model to the big data processes widens the scope of applicability of the model, making the model, generic. Section 4 discusses the development of 'custom big data stack' for each big data analytics application. This research paper concludes in Section 5 giving insights on the future research plans and possibilities.

## 2 BIG DATA LIFECYCLE
The big data lifecycle is composed of four main phases namely, data acquisition, storage, processing and visualization [14]. Data needs to be acquired from its respective sources and accumulated, filtered and pre-processed for analytics. This phase is commonly referred to as acquisition. Once data has been acquired, the next step involves the storage and processing of data to get the desired analytical results.

Finally, these analytics need to be presented to the user in a readable and usable format, which constitute the visualization phase. The different phases of the big data lifecycle have been illustrated in Fig. 1. Any big data analytics application can be modeled in the form of these four key processes to streamline the process of development.

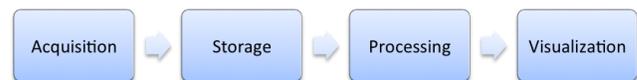

Fig. 1 – Big Data Lifecycle

## 3 PROPOSED MODEL
The proposed model characterizes the communication functions involved in a big data analytics system. In other words,

the proposed model partitions the big data analytics system into five abstract layers. As a rule, the layer lying below serves the layer lying above it, which is indicative of the flow of data in the system.

This proposed model is a conceptual model that can be used as a framework to fill in technologies on the basis of the proposed technology selection criteria for creation of technology stack. The three core layers of the model are Physical Layer, Data Storage and Processing Optimized Using Edge Computing (DSPOEC) Layer and Interaction Layer or Visualization Layer. The proposed model has been illustrated in Fig. 2.

The first layer of the model namely the physical layer consists of devices and sensors, which need to be connected to the cloud for data acquisition. The Cloud-Dew Architecture can be used for this purpose. It allows the user to use cloud services with each system controlling its own data and working seamlessly even in the absence of an Internet connection.

The Cloud-Dew Architecture is a computing system with two main features. The system includes independent systems like laptops, computers and mobile phones. These systems provide services on their own, which are referred to as micro-services. These services collaborate with the cloud services inherently. These systems have a web server installed on them, which is called Dew Server.

The Dew Server provides the services offered by the Cloud Server to these systems. Moreover, the Dew Server synchronizes the database of the system with that of the Cloud Server. Availability of dew servers is based on amount of uptime the device provides, battery backup, network latency and throughput, in addition to other parameters.

The Cloud-Dew architecture given by Wang [8] is shown in Fig. 3.

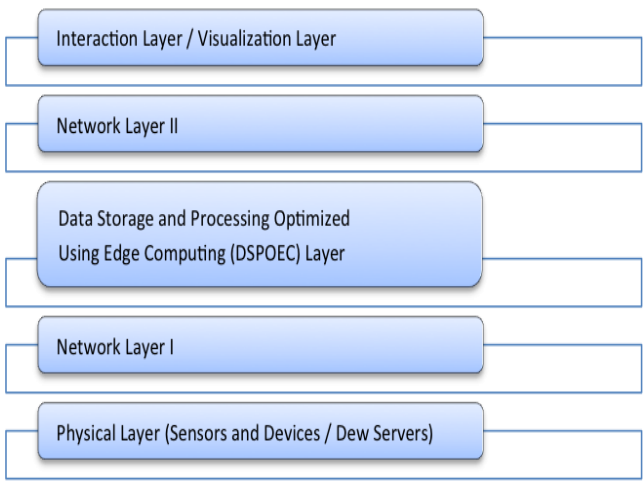

Fig. 2 – Proposed Model

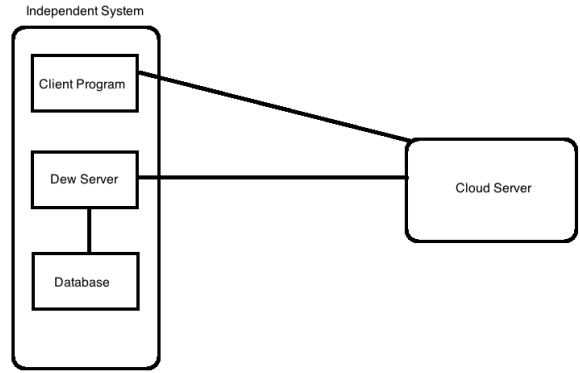

Fig. 3 – Cloud-Dew Architecture [8]

The Data Storage and Processing Optimized Using Edge Computing (DSPOEC) Layer will essentially be composed of cloud-based technologies for storage and processing of data (optimized in operation using edge computing principles). In edge computing, the processing ability of the network is pushed to the edges of the network. In other words, computing takes place closer to the devices.

Optimization of the system using Edge Computing concepts can considerably improve the performance of the system in addition to making the system more cost-effective. The network overhead of transferring data is reduced, improving the response time of the system.

Besides this, in applications that work on real-time, the data used for analysis becomes irrelevant rather quickly. The use of edge computing principles allows the system to discard irrelevant data, reducing infrastructure and storage costs.

Most of the big data tools are Hadoop-based, which is a big data processing platform [9]. Therefore, deployment of Hadoop on the cluster is inevitable. Machine learning tools can be used to assess the popularity of content and backhaul offloading for the implementation of proactive caching at the edge [10].

The 'Interaction layer or Visualization layer' will allow user intervention and interaction with the application. There are several cloud-based visualization tools like Pajek [11], Gephi [12] and VOSViewer [13] are available for big data analytics. On the basis of the requirements of the application, an appropriate tool can be integrated with the system.

The two network layers that are sandwiched between the three core layers facilitate communication of parameters and data from the preceding layer to the succeeding layer. They can also be seen as integration layers that allow technologies of one layer to interact with the technologies of the other layer. Evidently, the cloud supports all the five layers.

## 4 TECHNOLOGY STACK FOR CLOUD-BASED BIG DATA ANALYTICS

Technologies corresponding to the different layers of the proposed model need to be identified for each big data analytics application to create a 'custom big data stack'. The development of a big data stack for a field-specific application suffers from two fundamental challenges. Firstly, the data considered for big data analytics, for different applications, is different in nature, type and format. Secondly, data characteristics also vary on the basis of the field, application and data source. As a result, data model diversity and data characteristics variability are the two key challenges.

The existence of diverse data models is a probable solution to the first challenge. The taxonomy of the data models available for cloud-based platforms shall allow identification of an appropriate data model for the solution. Correspondingly, cloud-based data storage technologies can be chosen to solve this problem.

The requirements of an application with respect to a storage solution vary from one application to another. Hence, a decision of choosing a set of technologies is made by looking into the requirements and figuring out if the application requires indexing, support for concurrency and in-memory analytics.

These requirements are application-level requirements and once they are met, the next step is to look into cost, which majorly is a function of Storage or Network Overhead for big data. In order to choose the most optimal solution, required storage space needs to be approximated, which is where the concept of data model comes in.

With reference to network or processing overhead, performance testing of the proposed solution needs to be performed. Another important aspect of technology selection for Cloud-based big data analytics is the availability of the solution on the Cloud. Lastly, applications may also have other requirements like a need for an open-source solution. The selection criteria while choosing a storage solution for cloud-based big data analytics can be summarized as –

- The solution must meet the application-level requirements.
- The solution must meet the performance requirements.
- The solution must be Cloud-based.
- The solution must fulfill any other application-specific requirements.

Furthermore, the data characteristics cumulatively contribute towards the selection of computing model. Upon the identification of the computing model, the taxonomy of cloud-based technologies [14] with respect to computing models can be used for selecting technologies corresponding to the other phases of the big data lifecycle, which include data acquisition, processing and visualization. Technology selection from these two facets shall provide a 'custom big data stack' for the application concerned.

## 5  CONCLUSIONS

This research paper proposes a five-layer model for mapping the communication functions involved in a typical cloud-based big data analytics system. The layers of the proposed model are mapped onto the different phases of the big data lifecycle. The proposed approach performs technologies stack creation and selection criterion on the basis of the identified computing and data model for the application concerned.

The use of cloud as an infrastructural solution reduces the cost considerably. The cost-effectiveness of the proposed model makes it an ideal solution for SME's (Small and Medium-sized Enterprises) that can't opt for a private infrastructure and hardware for a sensor network.

Proposed future work in this area shall include benchmarking of technologies by creation of taxonomies for data and computing models and defining a clear selection criterion for technologies corresponding to each layer of the proposed model. Besides this, case studies to prove the viability of the proposed model shall also be performed.

ACKNOWLEDGMENTS
This work was supported by a grant from "Young Faculty Research Fellowship" under Visvesvaraya PhD Scheme for Electronics and IT, Department of Electronics & Information Technology (DeitY), Ministry of Communications & IT, Government of India.

REFERENCES
[1] Li, X., Song, J. and Huang, B., 2016. A scientific workflow management system architecture and its scheduling based on cloud service platform for manufacturing big data analytics. The International Journal of Advanced Manufacturing Technology, 84(1-4), pp.119-131.
[2] Baumann, P., Mazzetti, P., Ungar, J., Barbera, R., Barboni, D., Beccati, A., Bigagli, L., Boldrini, E., Bruno, R., Calanducci, A. and Campalani, P., 2016. Big data analytics for earth sciences: the earthserver approach. International Journal of Digital Earth, 9(1), pp.3-29.
[3] Pandey, M.K. and Subbiah, K., 2016, December. A Novel Storage Architecture for Facilitating Efficient Analytics of Health Informatics Big Data in Cloud. In Computer and Information Technology (CIT), 2016 IEEE International Conference on (pp. 578-585). IEEE.
[4] Hashem, I.A.T., Chang, V., Anuar, N.B., Adewole, K., Yaqoob, I., Gani, A., Ahmed, E. and Chiroma, H., 2016. The role of big data in smart city. International Journal of Information Management, 36(5), pp.748-758.
[5] Lewis, S.C. and Westlund, O., 2015. Big data and journalism: Epistemology, expertise, economics, and ethics. Digital Journalism, 3(3), pp.447-466.
[6] Sonka, S., 2014. Big data and the ag sector: More than lots of numbers. International Food and Agribusiness Management Review, 17(1), pp.1-20.
[7] Khan, S., Shakil, K.A. and Alam, M., 2016, December. Educational intelligence: Applying cloud-based big data analytics to the Indian education sector. In Contemporary Computing and Informatics (IC3I), 2016 2nd International Conference on (pp. 29-34). IEEE.
[8] Wang, Y., 2015. Cloud-dew architecture. International Journal of Cloud Computing, 4(3), pp.199-210.
[9] Patel, A.B., Birla, M. and Nair, U., 2012, December. Addressing big data problem using Hadoop and Map Reduce. In Engineering (NUiCONE), 2012 Nirma University International Conference on (pp. 1-5). IEEE.
[10] Zeydan, E., Bastug, E., Bennis, M., Kader, M.A., Karatepe, I.A., Er, A.S. and Debbah, M., 2016. Big data caching for networking: Moving from cloud to edge. IEEE Communications Magazine, 54(9), pp.36-42.
[11] De Nooy, W., Mrvar, A. and Batagelj, V., 2011. Exploratory social network analysis with Pajek (Vol. 27). Cambridge University Press.
[12] Bastian, M., Heymann, S. and Jacomy, M., 2009. Gephi: an open source software for exploring and manipulating networks. Icwsm, 8, pp.361-362.
[13] Van Eck, N.J. and Waltman, L., 2011. Text mining and visualization using VOSviewer. arXiv preprint arXiv:1109.2058.
[14] Khan, S., Shakil, K. A., Alam, M. 2017. Big Data Computing Using Cloud-Based Technologies: Challenges and Future Perspectives In: Networks of the Future: Architectures, Technologies, and Implementations. Chapman and Hall/CRC.
[15] Mansaf Alam, Kashish Ara Shakil, A decision matrix and monitoring based framework for infrastructure performance enhancement in a cloud based environment, Advances in Engineering and Technology Series, Volume 7, pp: 147-153, Elsevier.
[16] Alam, B., Doja, M. N., Alam, M., & Mongia, S. (2013). 5-Layered Architecture of Cloud Database Management System. AASRI Procedia, 5, 194-199.